\documentstyle[floats,aps,prd,eqsecnum,epsfig,amsfonts,amsmath]{revtex}

\newcommand{\gamo}{{\gamma_{\rm o}}}
\newcommand{\gamt}{{\gamma_{\rm t}}}
\newcommand{\Omn}{{\Omega_{\rm n}}}
\newcommand{\Omo}{{\Omega_{\rm o}}}
\newcommand{\Omt}{{\Omega_{\rm t}}}
\newcommand{\mun}{{\mu_{\rm n}}}
\newcommand{\muo}{{\mu_{\rm o}}}
\newcommand{\mut}{{\mu_{\rm t}}}
\newcommand{\sigm}{{\sigma_-}}
\newcommand{\sip}{{\sigma_+}}
\newcommand{\Sim}{\Sigma_-}
\newcommand{\Sip}{\Sigma_+}

\begin{document}

\title{Isotropization of two-component fluids}

\author{Martin Goliath\thanks{E-mail: goliath@physto.se}}
\address{Department of Physics, Stockholm University, 
  Box 6730, S--113 85 Stockholm, Sweden}
\author{Ulf S. Nilsson\thanks{E-mail: ulfn@physto.se}}
\address{Department of Applied Mathematics, University of Waterloo,
  Waterloo, Ontario N2L 3G1, Canada \\
  and \\ Department of Physics, Stockholm University, 
  Box 6730, S--113 85 Stockholm, Sweden}
\maketitle

\begin{abstract}
  We consider the problem of late-time isotropization in spatially
  homogeneous but anisotropic cosmological models when the source of
  the gravitational field consists of two non-interacting perfect
  fluids -- one tilted and one non-tilted. In particular, we study
  irrotational Bianchi type V models. By introducing appropriate
  dimensionless variables, a full global understanding of the state
  space of the gravitational field equations becomes possible. The
  issue of isotropization can then be addressed in a simple fashion. 
  We also discuss implications for the cosmic ``no-hair'' theorem for
  Bianchi models when part of the source is a tilted fluid.  
\end{abstract}


\section{Introduction}

Since the Bianchi models were introduced into cosmology \cite{Godel,Taub},
they have been the most studied generalizations of the spatially
homogeneous and isotropic Friedmann-Lema\^{\i}tre (FL) models. The
Bianchi models are spatially homogeneous but anisotropic, and thus
well suited for studying the effects of anisotropic expansion on the  
evolution of the universe. The complexity of a Bianchi model is
determined by its three-dimensional symmetry group $G_3$, in
conjunction with the chosen matter model. Since the present universe
seems to be very well described by a FL model, our main interest is
models that can be ``close to a FL model'' at late times. In this
paper we will restrict our consideration to models that may approach
the open FL model at late times. Within the class of Bianchi models,
only models of Bianchi type V and VII$_h$ can possibly have this
behavior (see Collins \& Hawking \cite{art:CollinsHawking1973a}). This
follows since the open FL metric admits a $G_3$ of these particular
Bianchi types. Here we will focus on the type V models. The notion of
a model approaching a FL model at late times is often referred to as
late-time isotropization. 

As regards the matter description, most studies of Bianchi cosmologies
use non-tilted barotropic perfect fluids with a linear equation of
state. The book by Wainwright \& Ellis \cite{book:WainwrightEllis1997}
is basically devoted to such models. In the so-called non-tilted
models, the fluid 4-velocity is orthogonal to the orbits of the
isometry group $G_3$. Models in which the 4-velocity of the fluid is
not orthogonal to the group orbits are referred to as tilted, and were
introduced by King \& Ellis \cite{art:CollinsEllis1979}. These models
have been studied by, for example, Collins \& Ellis
\cite{art:CollinsEllis1979}, Hewitt \& Wainwright
\cite{art:HewittWainwright1992}, and Harnett
\cite{thesis:Harnett1996}. Since the non-tilted models are a subset of
the tilted models, the latter can be viewed as simple generalizations
of the former. Another generalization of non-tilted models is achieved
by allowing the source of the gravitational field to be a combination
of two non-interacting, non-tilted perfect fluids. Since these models
can take into account both a radiation-dominated as well as a
matter-dominated epoch of the universe, they may be considered as more
physically relevant than single-fluid models. The qualitative behavior
of two-fluid models were studied by Coley \& Wainwright
\cite{art:ColeyWainwright1992} who showed that the models are
dominated at early times by the stiffer fluid, and at late times by
the softer fluid. This is the expected behavior of a universe
filled with radiation and dust. 

The next step is to allow one of the fluids in a two-fluid model to be
tilted, and subsequently, to let both fluids be tilted. Since the
dynamics of models containing tilted fluids is more difficult to
analyze than that of non-tilted models, we will consider the
combination of one non-tilted and one tilted fluid. The non-tilted
fluid is by definition irrotational, but tilted fluids can rotate. We
chose, however, to consider the subset of irrotational tilted
fluids (For a discussion of general single-fluid Bianchi type
V models, see Harnett \cite{thesis:Harnett1996}). We are thus
focusing on the subclass of irrotational Bianchi type V models. Both
fluids are assumed to satisfy linear barotropic equations of state, 
\begin{mathletters}
  \begin{eqnarray}
    p_{\rm o} &=& (\gamma_{\rm o}-1)\mu_{\rm o}\ , \\
    p_{\rm t} &=& (\gamma_{\rm t}-1)\mu_{\rm t}\ ,
  \end{eqnarray}
\end{mathletters}
where $p_{\rm o,t}$ are the pressures of the fluids, $\mu_{\rm o,t}$
the energy densities, and $\gamma_{\rm o,t}$ are constants that
satisfy $0\leq\gamma_{\rm o,t}\leq2$ with $\gamma_{\rm o}\neq\gamma_{\rm t}$.
We exclude, however, the specific values $2/3$ and $2$ since these
models behave qualitatively different. The indices ${\rm o}$ and
${\rm t}$ refer to the orthogonal and the tilted fluid respectively
and will be used throughout the paper. We will also assume that the
energy densities are non-negative, {\em i.e.}, 
\begin{mathletters}\label{eq:posmu}
  \begin{eqnarray}
    \mu_{\rm o} \geq 0\ , \\
    \mu_{\rm t} \geq 0\ .
  \end{eqnarray}
\end{mathletters}

To discuss isotropization in detail, we need a well-defined notion of
when a model is ``close to a FL model''. For models with a single
non-tilted fluid, a vanishing fluid shear defines the FL models (see,
for example, section 2.4 in \cite{book:WainwrightEllis1997}). It is,
however, not sufficient to demand that the fluid shear itself should
approach zero at late times, since this will occur in {\em any}
single-fluid Bianchi model, irrespectively of whether the model
isotropizes or not. As realized already by Kristian \& Sachs
\cite{art:KristianSachs1966}, the appropriate quantity to consider is
the dimensionless ratio formed by normalizing the fluid shear with the
fluid expansion, the so-called dimensionless shear.  It measures the 
dynamical importance of the shear compared to the expansion of the
fluid. Since there are now two fluids present, the notion of
isotropization at late times needs to be generalized as follows. We
say that a two-fluid model isotropizes at late times if the
dimensionless shear of {\em both} fluids vanish in this limit. This
issue was partially addressed by Goliath \& Ellis
\cite{art:GoliathEllis1999}, who considered models with a tilted fluid
and a non-zero cosmological constant. Such models are contained within
the models studied in this paper if we set $\gamma_{\rm o}$ equal to
zero. We will comment on aspects of the behavior of these models that
were not addressed in \cite{art:GoliathEllis1999}. 
In general, vanishing dimensionless shear is not sufficient for a
model to isotropize, since models can be Weyl-dominated in the future,
as is the case for Bianchi type VII$_0$ and Bianchi type VIII
models \cite{art:Wainwrightetal1999}. However, for Bianchi type V this
is not the case \cite{thesis:Harnett1996}. 

The plan of the paper is as follows. In Sec. \ref{sec:dynsys}, the
field equations are rewritten as a first-order system of autonomous 
ordinary differential equations. Reduced dimensionless variables are
then introduced leading to a compact state space. In Sec. \ref{sec:local}
we perform a local analysis of the system of equations.
In Sec. \ref{sec:orth}, Bianchi type I and type V models with two
orthogonal fluids are studied, while the locally rotationally
symmetric type V models are considered in Sec. \ref{sec:LRS}. The
question of isotropization is discussed in detail. We end with a
discussion in Sec. \ref{sec:conc}. In App. \ref{app:kinprop}, the
kinematical properties of both fluids are given. The relationship to
the variables used in \cite{art:ColeyWainwright1992} is discussed in
App. \ref{app:chi}. We use units such that $c=8\pi G=1$. Orthonormal
frame indices are denoted by latin letters $a,b,...$\ .

\section{The gravitational field equations}\label{sec:dynsys}

The line element for the irrotational Bianchi type V models can be
written
\begin{equation}\label{eq:ds2}
  ds^2 = -dt^2 + D_1(t)^2dx^2 + {\rm e}^{-2x}\left[D_2(t)^2dy^2 +
  D_3(t)^2dz^2\right]\ ,
\end{equation}
(see, for example, Ellis \& MacCallum \cite{art:EllisMacCallum1969}).
In the models we consider there are two preferred timelike
congruences.  First we have the congruence associated with the normal
to the spatial symmetry surfaces, which is also, by definition, the
congruence of the 4-velocity of the non-tilted fluid, $u^a_{\rm
o}$. The time variable $t$ in (\ref{eq:ds2}) is chosen so that the
normal to the symmetry surfaces is $\tfrac{\partial}{\partial t}$. 
The other preferred congruence is that of the tilted fluid. The specific
form of the line element, Eq.\ (\ref{eq:ds2}), guarantees that this
fluid is irrotational. Hence, the 4-velocity of this fluid, $u_{\rm t}^a$, is
constrained by the field equations to be of a particular form, and can
conveniently be parameterized in terms of the so-called
{\em tilt variable} $v$ according to 
\begin{equation}
  u_{\rm t}^a = \Gamma\left(1,v,0,0\right)\ , \quad
  \Gamma = (1-v^2)^{-1/2}\ ,
\end{equation}
where the value $v=0$ corresponds to a non-tilted fluid.
The energy-momentum tensor of each fluid is 
\begin{mathletters}
  \begin{eqnarray}
    T_{\rm o}^{ab}&=&\left[\gamo u_{\rm o}^au_{\rm o}^b +
    (\gamo-1)\eta^{ab} \right]\muo \ , \\
    T_{\rm t}^{ab}&=&\left[\gamt u_{\rm t}^au_{\rm t}^b +
    (\gamt-1)\eta^{ab} \right]\mut \ .
  \end{eqnarray}
\end{mathletters}
The assumption that the two fluids are non-interacting leads to equations
of motion of the form 
\begin{equation}\label{eq:fluideq}
  \nabla_aT^{ab}_{\rm o} = 0 = \nabla_aT^{ab}_{\rm t} \ .
\end{equation}
We choose to parameterize the gravitational field using the expansion,
$\theta$, and the non-vanishing components of the shear-tensor
$\sigma_\pm$, of the normal congruence (and thus of the non-tilted
fluid). They are given by 
\begin{equation}
  \theta = \frac{d}{dt}\ln\left(D_1D_2D_3\right) \ , \quad
  \sip=-\frac{1}{2}\frac{d}{dt}\ln\left(\frac{D_1^2}{D_2D_3}\right) \ ,
  \quad 
  \sigm=\frac{\sqrt{3}}{2}\frac{d}{dt}\ln\left(\frac{D_2}{D_3}\right)
 \ .
\end{equation}
The expansion and shear of the tilted fluid can be written as
functions of these variables, see App. A. The gravitational field
equations,
\begin{equation}
  G_{ab} = T^{ab}_{\rm o} + T^{ab}_{\rm t}\ ,
\end{equation}
and the equations of motion of the fluids, Eq.\ (\ref{eq:fluideq}),
become 
\begin{mathletters}\label{eq:dimfuleq}
  \begin{center}
    {\bf Evolution equations}
  \end{center}
  \begin{eqnarray}
    \dot{\theta} &=& -\tfrac13\theta^2 - \tfrac{2}{3}\left(\sip^2 +
    \sigm^2\right) - \tfrac12(3\gamo-2)\muo - \frac12\frac{(3\gamt-2) +
      (2-\gamt)v^2}{1+(\gamt-1)v^2}\mun\ , \\ 
    \dot{\sigma}_+ &=& -\left(\theta - 2vB_1\right)\sip \ , \\
    \dot{\sigma}_- &=& -\theta\sigm \ , \\
    \dot{B}_1 &=& -\tfrac13\left(\theta-2\sip\right)B_1 \ , \\
    \dot{v} &=&
    \frac{v(1-v^2)}{3\left[1-(\gamt-1)v^2\right]}\left[2\sip
    +(3\gamt-4)\theta - 6(\gamt-1)B_1v\right]\ , \\
    \dot{\mu}_{\rm o} &=& -\gamo\theta\muo \ .
  \end{eqnarray}
  \begin{center}
    {\bf Constraint equation}
  \end{center}
  \begin{eqnarray}
    0 &=& \gamt v\mun + 2(1+(\gamt-1)v^2)B_1\sip \ .
  \end{eqnarray}
  \begin{center}
    {\bf Defining equation for $\mun$}
  \end{center}
  \begin{eqnarray}\label{eq:mu2def}
    3\mun &=& \theta^2 - \sip^2 - \sigm^2 - 9B_1^2 - 3\muo \ ,
  \end{eqnarray}
\end{mathletters}
where we have introduced 
\begin{mathletters}
  \begin{eqnarray}
    B_1 &=& D_1^{-1}\ , \\
    \mun &=& \frac{1+(\gamt-1)v^2}{1-v^2}\mut \ , \label{eq:mundef}
  \end{eqnarray}
\end{mathletters}
and a dot denotes differentiation with respect to $t$. 

From the form of Eq.\ (\ref{eq:mu2def}), it is now clear why our
choice of variables is a good one. The assumption in Eq.\ (\ref{eq:posmu}),
in conjunction with Eq.\ (\ref{eq:mu2def}) and Eq.\ (\ref{eq:mundef}),
implies that $\theta$ cannot change sign. Therefore we can, without
loss of generality, assume that $\theta\geq 0$, {\em i.e.}, we
restrict ourselves to expanding models. From Eq.\ (\ref{eq:mu2def}) it
also follows that $\theta$ is a dominant quantity. We then
introduce bounded dimensionless ``$\theta$-normalized'' variables for
which the system of equations (\ref{eq:dimfuleq}), is reduced as far as
possible, and for which the state 
space is compact. These variables are defined by
\begin{equation}
  \Sigma_\pm=\frac{\sigma_\pm}{\theta}, \quad
  A=\frac{3B_1}{\theta}, \quad
  \Omo=\frac{3\muo}{\theta^2}, \quad 
  \Omn=\frac{3\mun}{\theta^2} \ .
\end{equation}
The subsequent introduction of a dimensionless time variable $\tau$,
which satisfies
\begin{equation}
  \frac{d\tau}{dt} = \frac{\theta}{3}\ ,
\end{equation}
leads to a decoupling of the $\theta$-equation, which can be written on
the form
\begin{equation}\label{eq:q}
  \theta^\prime = -(1+q)\theta \ , \quad
  q=2-2A^2-\frac{3(2-\gamo)}{2}\Omo-\frac{3(2-\gamt)+(5\gamt-6)v^2} 
  {2\left[1+(\gamt-1)v^2\right]}\Omn ,
\end{equation}
where the prime denotes differentiation with respect to $\tau$. The 
parameter $q$ is the deceleration parameter associated with the normal
congruence and the non-tilted fluid. The remaining equations
can now be written on dimensionless form as follows.
\begin{mathletters}\label{eq:dynsyst}
  \begin{center}
    {\bf Evolution equations}  
  \end{center}
  \begin{eqnarray}
    \Sip^\prime&=&-(2-q-2Av)\Sip , \\
    \Sim^\prime&=&-(2-q)\Sim \ , \\
    A^\prime&=&(q+2\Sip)A , \\
    v^\prime&=&\frac{v(1\!-\!v^2)}{1-(\gamt\!-\!1)v^2}
    \left[2\Sip+3\gamt-4 -2(\gamt\!-\!1)Av\right] \ , \\
      \Omo' &=& \left[2q-(3\gamo-2)\right]\Omo\ .
  \end{eqnarray}
  \begin{center}
    {\bf Constraint equation}
  \end{center}
  \begin{eqnarray}\label{eq:constraint}
    0 &=& \gamt v\Omn+2[1+(\gamt-1)v^2]A\Sip \ .
  \end{eqnarray}
  \begin{center}
    {\bf Defining equation for $\Omn$}
  \end{center}
  \begin{eqnarray}
    \Omn &=& 1-\Sip^2 - \Sim^2 - A^2 - \Omo\ .
  \end{eqnarray}
\end{mathletters}

The set of equations (\ref{eq:dynsyst}) shows that the irrotational
Bianchi type V models with one orthogonal and one tilted fluid is
governed by a system of five autonomous ordinary differential
equations subject to one constraint.  The dimension of
the state space is thus four. The set (\ref{eq:dynsyst}) is
invariant under the discrete transformations
\begin{equation}
  \left(\Sip, \Sim,  A,  v, \Omo\right) \rightarrow
  \left(\Sip, \Sim, -A, -v, \Omo\right)\ , \quad 
  \left(\Sip,  \Sim, A, v, \Omo\right) \rightarrow
  \left(\Sip, -\Sim, A, v, \Omo\right)\ ,
\end{equation}
so we can, without loss of generality, restrict ourselves to
the invariant set defined by $A\geq0$ and $\Sim\geq 0$. The variables
$\Sip, \Sim, A, v,$ and $\Omo$ therefore satisfy $0\leq
\Sip^2,\Sim,A,v^2, \Omo \leq 1$. There is a number of important
subsets of Eqs.\ (\ref{eq:dynsyst}):

\begin{enumerate}
  \item  The three-dimensional subset defined by $v=0$, which describes
  a Bianchi type V universe with two non-tilted fluids. This case is
  thus included in the general study of Coley \& Wainwright
  \cite{art:ColeyWainwright1992}. We will, however, consider these
  models in Sec. \ref{sec:orth} for the purpose of comparison with the
  models where one fluid is tilted. The subset $v=0$  also contains
  Bianchi type I models and the open FL model as special cases. 
  \item The three-dimensional subset $\Sim=0$, which corresponds to
  locally rotationally symmetric (LRS) models (see, for example,
  \cite{book:ExactSolutions1980}). This subset turns out to be very important
  for the evolution of the general irrotational type V models and
  will be considered in detail in Sec. \ref{sec:LRS}.  
\end{enumerate}

The next step in the analysis is to consider the equilibrium points of
Eq.\ (\ref{eq:dynsyst}). This is done in the next section.

\section{Qualitative analysis}\label{sec:local}

\begin{center}
  \begin{table}
    \begin{tabular}{l|ccccc|l}
      & $\Sip$ & $\Sim$ & $A$ & $v$ & $\Omo$ & Restrictions \\ \hline
      $F^0_{\rm t}$ & 0 & 0 & 0 & 0 & 0 & \\
      $F^0_{\rm o}$ & 0 & 0 & 0 & 0 & 1 & \\
      $M^0$ & 0 & 0 & 1 & 0 & 0 & \\
      $K^0$ & \multicolumn{2}{c}{$\Sip^2 + \Sim^2 = 1$} & 0 & 0 & 0 & \\
      $\tilde{M}$ & 0 & 0 & 1 & $v = \tfrac{3\gamt-4}{2(\gamt-1)}$ & 0 & 
      $\tfrac{6}{5}<\gamt<2$ \\ 
      ${\cal C}^\pm$ & $\Sip = -\tfrac12(3\gamt-4)$ &
      $\Sim = \pm\tfrac12\sqrt{3(2-\gamt)(3\gamt-2)}$ & 0 & $v$ & 0 & \\
      $F^v_{\rm o}$ & 0 & 0 & 0 & $v$ & 1 & $\gamt = \tfrac{4}{3}$ \\
      $F^\pm_{\rm o}$ & 0 & 0 & 0 $\pm1$ & 1 & \\
      $M^\pm$ & 0 & 0 & 1 & $\pm1$ & 0 & \\
      $K^\pm$ & \multicolumn{2}{c}{$\Sip^2 + \Sim^2 = 1$} & 0 & $\pm1$
      & 0 & \\
      {$\cal H$} & $\Sip = -1+A$ & 0 & $A$ & 1 & 0 & \\ 
    \end{tabular}
    \caption{Equilibrium points for the irrotational Bianchi type V
      models with one orthogonal and one tilted perfect fluid.
      Unless indicated, all equilibrium points and sets are in
      the physical part of state space for
      $0\leq\gamma_{\rm o,t}<2$, $\gamma_{\rm o,t}\neq2/3$.}\label{tab:equi} 
  \end{table}
\end{center}

In Table \ref{tab:equi}, we present the equilibrium points of the system
(\ref{eq:dynsyst}). The corresponding eigenvalues are given in
Table \ref{tab:eigen}. Since the constraint,
Eq.\ (\ref{eq:constraint}), cannot be solved globally in
an analytic way, it will be treated locally (see, for example,
\cite{art:HewittWainwright1992}). 

Some of the equilibrium points correspond to exact solutions of the
field equations. For example, the equilibrium points $F^0_{\rm t}$ and
$F^0_{\rm o}$ correspond to flat FL models in which the tilted fluid
and the non-tilted fluid is dominant, respectively (The
``tilted'' fluid is in fact non-tilted at $F^0_{\rm t}$ since $v=0$,
but we refer to it as the tilted fluid for simplicity). The point $M^0$  
is the Milne model, while the equilibrium set $K^0$ corresponds to
Kasner-like models. The point $\tilde{M}$ corresponds to flat space
and coincides with $M^0$ for $\gamt=4/3$. We also note
the appearance of the equilibrium set $F^v_{\rm o}$ for the specific
value $\gamt=\tfrac{4}{3}$. It is associated with a line bifurcation  
that transfers stability between the equilibrium points
$F^\pm_{\rm o}$ and  $F^0_{\rm o}$. The points ${\cal C}^\pm$ correspond
to particular Kasner solutions. There is also a number of equilibrium
points for which the tilt is extreme ($v^2=1$). Whether these
equilibrium points correspond to exact Bianchi solutions or not seems
to be an open question (see comment on p. 4245 in
\cite{art:HewittWainwright1992}). We note that the constraint,
Eq.\ (\ref{eq:constraint}), is degenerate ($\nabla G=0$) at
the point $F_{\rm o}^0$, allowing all five eigenvector directions to
be physical at this point. 

\begin{center}
  \begin{table}
    \begin{tabular}{l|ccccc|c}
      & \multicolumn{5}{c|}{Eigenvalues} & Elim. \\ \hline
      
      $F^0_{\rm t}$ &
      $-\tfrac{3}{2}(2-\gamt)$ &
      $-\tfrac{3}{2}(2-\gamt)$ &
      $\tfrac{1}{2}(3\gamt-2)$ &
      $3(\gamt - \gamo)$ &
      & $v$ \\
      
      $F^0_{\rm o}$ &
      $-\tfrac{3}{2}(2-\gamo)$ &
      $-\tfrac{3}{2}(2-\gamo)$ &
      $\tfrac{1}{2}(3\gamo-2)$ &
      $3\gamt-4$ &
      $-3(\gamt-\gamo)$ &
      -- \\
      
      $M^0$ &
      $-2$ &
      $-(3\gamt-2)$ &
      $3\gamt-4$ &
      $-(3\gamo-2)$ &
      & $\Sip$ \\

      $K^0$ &
      0 &
      $3(2-\gamt)$ &
      $2\Sip + 3\gamt-4$ & 
      $3(2-\gamo)$ &
      & $A$ \\
      
      $\tilde{M}$ &
      $-2$ &
      $-\tfrac{2-\gamt}{\gamt-1}$ &
      $-\tfrac{(3\gamt-4)(5\gamt-6)}{(\gamt-1)(9\gamt-10)} $ &
      $-(3\gamo-2)$ &
      & $\Sip$ \\

      ${\cal C}^\pm$ &
      $3(2-\gamt)$ &
      0 &
      0 &
      $3(2-\gamo)$ &
      & $A$ \\ 

      $F^v_{\rm o}$ & 
      $-\tfrac{3}{2}(2-\gamo)$ &
      $-\tfrac{3}{2}(2-\gamo)$ &
      $\tfrac{1}{2}(3\gamo-2)$ &
      0 &
      & $\Omo$ \\

      $F^\pm_{\rm o}$ &
      $-\tfrac{3}{2}(2-\gamo)$ &
      $-\tfrac{3}{2}(2-\gamo)$ &
      $\tfrac{1}{2}(3\gamo-2)$ &
      $-\tfrac{2(3\gamt-4)}{2-\gamt}$ &
      & $\Omo$ \\

      $M^+$ &
      $-2$ &
      0 &
      2 &
      $-(3\gamo-2)$ &
      & $\Sip$ \\
      
      $M^-$ &
      $-2$ &
      $-4$ &
      $-\tfrac{2(5\gamt-6)}{2-\gamt}$ &
      $-(3\gamo-2)$ &
      & $\Sip$ \\

      $K^\pm$ &
      $2(1+\Sip)$ &
      0 &
      $-\tfrac{2}{2-\gamt}\left[2\Sip + (3\gamt-4)\right]$ &
      $3(2-\gamo)$ &
      & $A$ \\

      ${\cal H}$ & 
      0 &
      $-2(1+\Sip)$ &
      $2(1-2\Sip)$ &
      $-(4\Sip + 3\gamo-2)$ &
      & $A$ \\
    \end{tabular}
    \caption{Eigenvalues for the equilibrium points of the
      irrotational Bianchi V models, with one orthogonal and one
      tilted perfect fluid.}\label{tab:eigen} 
  \end{table}
\end{center}

\section{Two orthogonal fluids}\label{sec:orth}

\begin{figure}
  \begin{minipage}[b]{.45\linewidth}
    \centering\epsfig{file=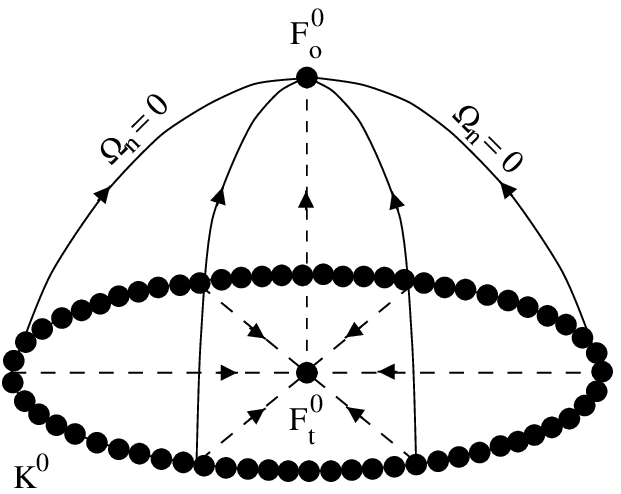, width=0.7\textwidth}
    \caption{The state space of two-fluid orthogonal Bianchi type I
      models with the choice $\gamt>\gamo$. The bottom and the
      ``dome'' correspond to one-fluid models.}\label{fig:orth-A0}
  \end{minipage}\hfill
  \begin{minipage}[b]{.45\linewidth}
    \centering\epsfig{file=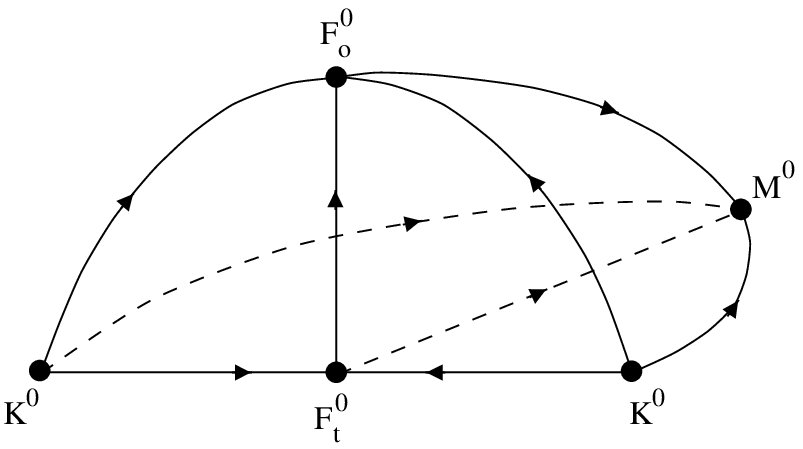, width=0.9\textwidth}
    \caption{The state space of two-fluid orthogonal Bianchi type V
    models with the choice $\gamt>\gamo$. The bottom and ``dome''
    correspond to type V one-fluid models.}\label{fig:orth-Sip0}
  \end{minipage}\hfill
\end{figure}

Setting $v=0$ in the constraint Eq.\ (\ref{eq:constraint}), implies
either $A=0, \Sip\neq0$ (Bianchi type I models) or $\Sip=0, A\neq0$
(Bianchi type V models). Without loss of generality we can assume that
$\gamt>\gamo$. The boundary subsets of the state space for these two
classes of models are given by the two invariant submanifolds $\Omo=0$
and $\Omn=0$, which describe the corresponding one-fluid models. The
dynamics of the Bianchi I state space is shown in Fig.\
\ref{fig:orth-A0}, while the dynamics of the type V models is shown in
Fig.\ \ref{fig:orth-Sip0}.

For type I models there are two sources. The equilibrium point
$F_{\rm t}^0$ gives rise to a single orbit ending at $F_{\rm o}^0$. It
corresponds to a flat Friedmann model with two orthogonal
fluids. The other source is the equilibrium set $K^0$, of which each
point is associated with a one-parameter set of orbits. Therefore
this equilibrium set describes the generic behavior at early times.
The future attractor of these orbits is the point $F^0_{\rm o}$. Thus,
from Eqs.\ (\ref{eq:sio}) and (\ref{eq:sito}), all orthogonal two-fluid
Bianchi type I models isotropize. 

For type V models there are three sources. The equilibrium point
$F_{\rm t}^0$ is associated with a one-parameter set of orbits
(characterized by $\Sim=0$), which corresponds to open Friedmann
models with two orthogonal fluids. They all end at the equilibrium
point $M^0$. The other two sources are the two equilibrium points
belonging to the set $K^0$ with $\Sim=\pm1$. Both of these are
associated with two-parameter sets of orbits, and thus describe the
generic early-time behavior. They are future attracted to
$M^0$. Consequently, from Eqs.\ (\ref{eq:sio}) and (\ref{eq:sito}),
all orthogonal two-fluid Bianchi type V models isotropize. 

From the analysis of the orthogonal two-fluid models, it is clear that
the generic behavior is described by equilibrium points associated
with two-parameter sets. Thus, in what follows, we will focus on such
equilibrium points. Note that the variable $\chi$ of Coley \&
Wainwright \cite{art:ColeyWainwright1992} (see App. \ref{app:chi}) is a
monotone function when both the fluids are orthogonal. This is no
longer the case when one of the fluids is tilted.

\noindent

\section{Bianchi type V LRS models with one tilted and one orthogonal
  fluid}\label{sec:LRS}

As Bianchi type I models do not allow the combination of one
non-tilted and one tilted fluid as a source due to the constraint
Eq.\ (\ref{eq:constraint}), we focus on the Bianchi type V models. Since
$\Omo,\Omn>0$ imply $q<2$ by Eq.\ (\ref{eq:q}), the evolution equation for
$\Sim$ implies that $\Sim$ is a monotone decreasing function along all
orbits with $\Sim\geq0$ and $\Omo,\Omn >0$. This fact significantly
restricts the evolution at late times. It implies that
$\lim_{\tau\rightarrow\infty} \Sim = 0$, 
for all orbits with $\Omo,\Omn > 0$. This can be proven along the
lines used to prove the similar statement for tilted single-fluid
models, see \cite{art:HewittWainwright1992}. The asymptotic behavior
as $\tau\rightarrow\infty$ is thus contained in the three-dimensional
invariant set $\Sim=0$ corresponding to the LRS models.

The three Kasner circles $K^0$ and $K^\pm$ each reduce to two 
equilibrium points in the LRS submanifold, namely points for which
$\Sip=\pm1$. We denote these equilibrium points $K^0_\pm$, $K^+_\pm$,
and $K^-_\pm$, where the subscript distinguishes between the two signs
of $\Sip$. The equilibrium points for which $v$ and $\Sip$ have the
same sign (collectively denoted $K^\pm_\pm$) are not located on the
boundary of the interior of the LRS submanifold. Consequently, we do
not need to consider them when studying the dynamics. 

\noindent
\begin{figure}[t]
  \begin{minipage}[b]{.45\linewidth}
    \centering\epsfig{file=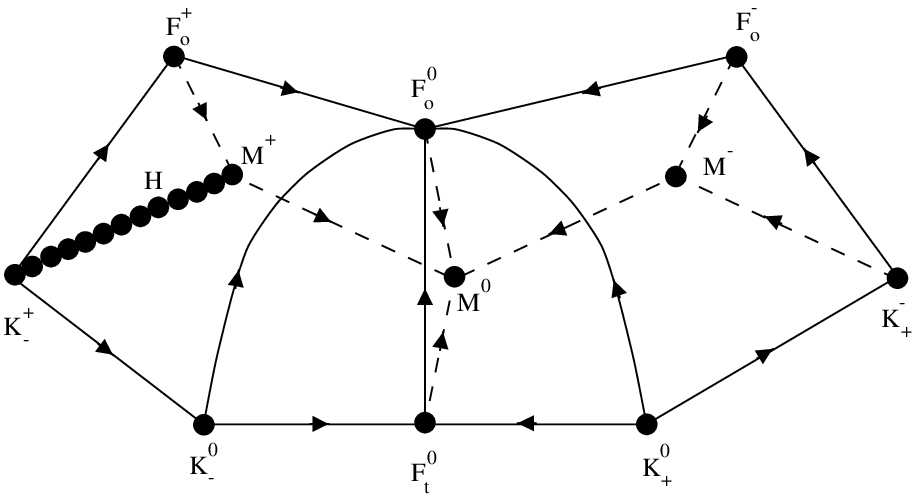, width=0.9\textwidth}
    \caption{The state space of Bianchi type V LRS models when
      with $2/3<\gamo<\gamt\leq6/5$.}\label{fig:lrs-1} 
  \end{minipage}\hfill
  \begin{minipage}[b]{.45\linewidth}
    \centering\epsfig{file=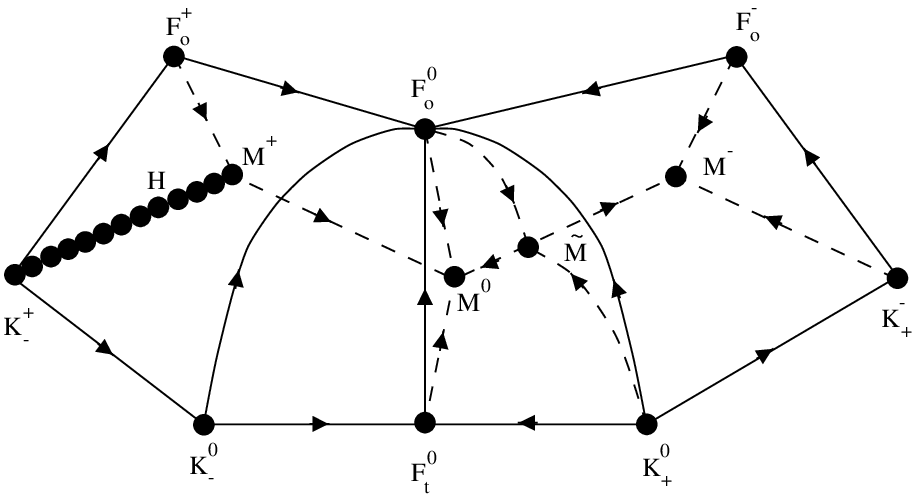, width=0.9\textwidth}
    \caption{The state space of Bianchi type V LRS models submanifold
      when $\gamt>\gamo$ with $2/3<\gamo<4/3$ and
      $6/5<\gamt<4/3$.}\label{fig:lrs-2} 
  \end{minipage}\hfill
  \begin{minipage}[b]{.45\linewidth}
    \centering\epsfig{file=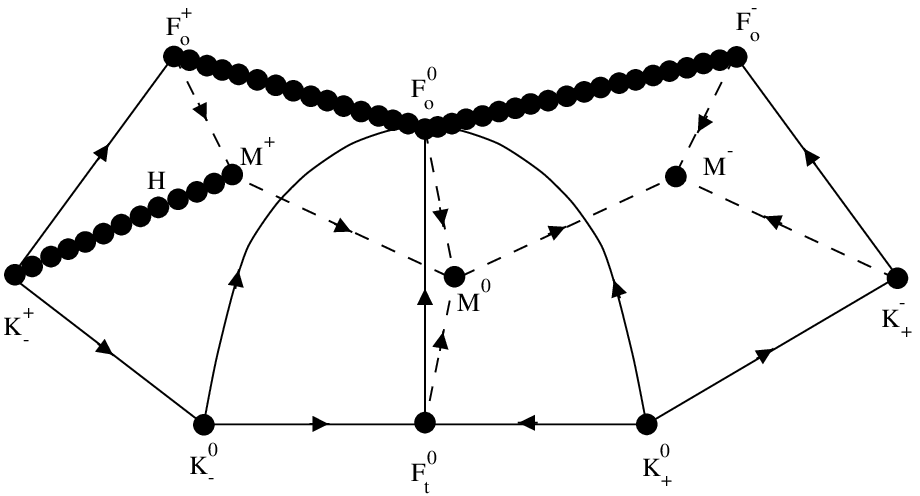, width=0.9\textwidth}
    \caption{The state space of Bianchi type V LRS models when
      $\gamt=4/3$ and $2/3<\gamo<4/3$}\label{fig:lrs-3a}
  \end{minipage}\hfill
  \begin{minipage}[b]{.45\linewidth}
    \centering\epsfig{file=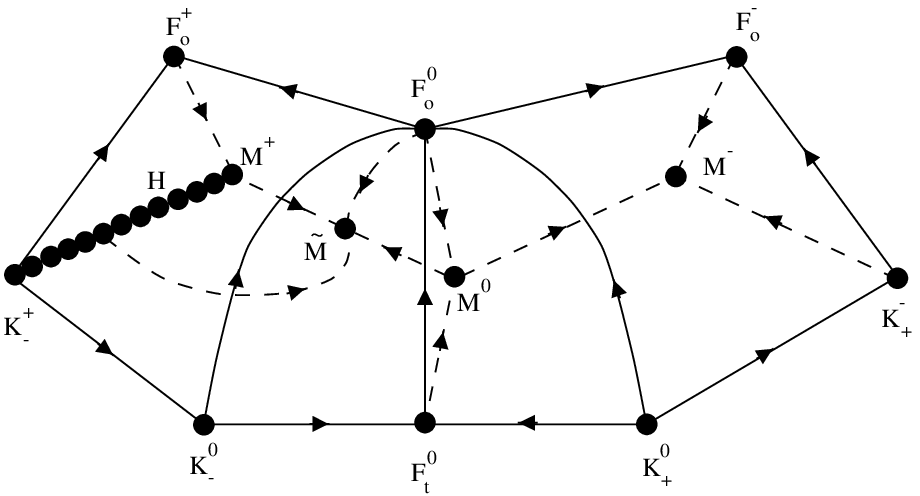, width=0.9\textwidth}
    \caption{The state space of Bianchi type V LRS models when
      $\gamt>\gamo$ with $2/3<\gamo<2$ and
      $4/3<\gamt<2$.}\label{fig:lrs-3} 
  \end{minipage}\hfill
\end{figure}

The state space for the LRS submanifold with $2/3<\gamo<\gamt<2$ is
presented in Figs. \ref{fig:lrs-1}--\ref{fig:lrs-3}. The effect of
changing the equation-of-state parameters so that $\gamo>\gamt$ is
that the flow along the orbit between $F^0_{\rm o}$ and $F^0_{\rm t}$ is
reversed. Similarly, $0\leq\gamo<2/3$ results in a stability change along
the orbits from the equilibrium points $F^*_{\rm o}$ to $M^*$, where
$*\in\{0,+,-\}$. Note that the special case $\gamo=0$ corresponds to a
cosmological constant. Consequently, the LRS state space for
$0\leq\gamo<2/3$ is contained in Figs. 9--11 of
\cite{art:GoliathEllis1999}. Finally, if $0\leq\gamt<2/3$ the flow
changes along the orbits $F^0_t$--$M^0$ and $K^0_+$--$K^-_+$.

The sources and sinks for various equations of state are summarized in 
Table \ref{tab:lrs-stab}. The isotropization properties, which can be
found from Eqs.\ (\ref{eq:sio}) and (\ref{eq:sit}), are also listed.
For $\gamt<6/5$, all models isotropize. for $\gamo>2/3$,
$6/5<\gamt<4/3$ there is a class of solutions of non-zero measure that
does not isotropize, namely the class of solutions associated with
$M^-$. For $\gamo>2/3$, $\gamt=4/3$, all models isotropize. For
$\gamt>4/3$ no models isotropize. Note that the physically interesting
combination of one dust fluid and one radiation fluid {\em always}
isotropizes, regardless of which fluid is tilted. 

When $\gamo<2/3,\gamt>4/3$, the future attractors are
$F_{\rm o}^\pm$. Solutions corresponding to orbits approaching these
equilibrium points do not isotropize, see Table \ref{tab:lrs-stab},
even though the final states are inflationary in the sense that the
deceleration parameter associated with the normal congruence
$q=\tfrac12(3\gamo-2)$ is negative. Thus, in some sense it seems to be
misleading to refer to solutions corresponding to orbits approaching
these equilibrium points as ``asymptotically Friedmann models'',
although the solutions corresponding to the points themselves are
Friedmann models. In particular, for the case of a cosmological constant 
$\gamo=0$, the solutions corresponding to $F_{\rm o}^\pm$ are de
Sitter models \cite{art:GoliathEllis1999}. This is consistent with
the cosmic ``no-hair'' theorem for Bianchi models \cite{art:Wald1983}.
However, the theorem does not guarantee that the tilt tends to zero, as
pointed out by Raychaudhuri \& Modak \cite{art:RaychaudhuriModak1988}.
From our analysis, it is clear that this cautionary note is crucial
for these models. As the tilted fluid does {\em not} become orthogonal
at late times, the expansion-normalized shear of the tilted fluid does
not vanish at late times. This stresses the point made in
\cite{art:GoliathEllis1999} that one should be cautious about the
isotropization of tilted models. As seen in Table \ref{tab:lrs-stab}, this
is in fact the generic behavior for models with $0\leq\gamo<2/3$ and 
$\gamt\geq4/3$.  

For general irrotational Bianchi type V models, there is no
{\em a priori} reason that $\Sigma_{{\rm t}-}\rightarrow0$ when
$\Sigma_{{\rm t}+}\rightarrow0$. However, this is indeed the case for
the equilibrium points in question, as can be seen from Eq.\
(\ref{eq:Sigma-fluid}), noting that $\Sigma_-\rightarrow0$. Thus, the
analysis of isotropization for the LRS submanifold holds for the
general class of models as well.

\section{Discussion}\label{sec:conc}

In this paper we have continued the study of irrotational Bianchi type V
cosmologies, using the dynamical systems approach initiated by Hewitt
\& Wainwright \cite{art:HewittWainwright1992} and Coley \&
Wainwright \cite{art:ColeyWainwright1992}. The source of the
gravitational field has been taken to be two non-interacting fluids, one
orthogonal and one tilted. Such models can describe a universe where
one of the fluids models the contribution of radiation to the
energy density of the universe, and the other the matter content.

We have found that, although the orthogonal two-fluid models
isotropize, this is not necessarily the case when one of the fluids is
tilted. Thus, depending on the equation-of-state parameters $\gamo$
and $\gamt$, it is possible to find cases for which all or a subset of
the solutions are anisotropic to the future. In particular, there 
are models which are inflationary, but do not isotropize. However, we
emphasize that the cases of dust plus radiation always isotropize. 

It should be stressed that the relevant quantities when determining
whether the shear dies away are shear quantities normalized with
respect to the expansion associated with each fluid.

\section*{Acknowledgements}

We thank Claes Uggla for useful comments.
MG was supported by a grant from C F Liljevalch J:ors
stipendiefond. USN was supported by G{\aa}l\"ostiftelsen,
Svenska Institutet, Stiftelsen Blanceflor and the
University of Waterloo.

\begin{table}
  \begin{tabular}{c|cc|cc}
    & \multicolumn{4}{c}{Sources} \\ \hline
    $0\leq\gamt<2/3$ & \multicolumn{2}{c}{$K^-_+$, ${\cal H}$} \\
    $ 2/3<\gamt<2$   & \multicolumn{2}{c}{$K^0_+$, ${\cal H}$} \\\hline\hline
    & \multicolumn{4}{c}{Sinks} \\
    & \multicolumn{2}{c}{$0\leq\gamo<2/3$} &
      \multicolumn{2}{c}{$2/3<\gamo<2$} \\ \hline
      $0\leq\gamt<2/3$ & $F^0_{\rm t}$ ($\gamt<\gamo$) & all isotropize &
    $F^0_{\rm t}$ & all isotropize \\
    & $F^0_{\rm o}$ ($\gamt>\gamo$) & \\
    $2/3<\gamt<6/5$  & $F^0_{\rm o}$ & all isotropize &
    $M^0$ & all isotropize \\
    $6/5<\gamt<4/3$  & $F^0_{\rm o}$ & all isotropize &
    $M^0$, $M^-$ & $M^0$ isotropize \\
    $\gamt=4/3$      & $F^v_{\rm o}$ & none isotropize (except for $v=0$) &
    $M^0$, $M^-$ & all isotropize \\ 
    $4/3<\gamt<2$    & $F^\pm_{\rm o}$ & none isotropize &
    $\tilde{M}$, $M^-$ & none isotropize \\
  \end{tabular}
  \caption{Sources and sinks for the Bianchi type V LRS models with
    two fluids.}\label{tab:lrs-stab} 
\end{table}

\appendix

\section{Fluid properties}\label{app:kinprop}

Here we present the kinematical properties of the two fluids in terms
of the variables used to parameterize the gravitational field. For the
non-tilted fluid we have\\
\mbox{}\\
{\bf Fluid expansion}
\begin{equation}
  \theta_{\rm o} = \theta\ ,
\end{equation}
{\bf Fluid shear}
\begin{equation}
  {\sigma_{\rm o}}^2 = \sip^2 + \sigm^2
\end{equation}
The dimensionless shear $\Sigma_{\rm o}$ of the non-tilted fluid is thus
\begin{equation}\label{eq:sio}
  \Sigma_{\rm o}^2= \frac{{\sigma_{\rm o}}^2}{\theta_{\rm o}^2}
  =\Sip^2 + \Sim^2 .
\end{equation}
The fluid properties of the tilted fluid are as follows.\\
\mbox{}\\
{\bf Fluid expansion}
\begin{equation}
  \theta_{\rm t} = \frac{1}{\sqrt{1-v^2}}\left[ \frac{3\theta - 6vB_1
  + v^2(2\sip - \theta)}{3\left[1-(\gamt-1)v^2\right]}\right] = 
  \frac{1}{\sqrt{1-v^2}}\left[ \frac{3 - 2vA
  + v^2(2\Sip - 1)}{3\left[1-(\gamt-1)v^2\right]}\right]\theta\ ,
\end{equation}
{\bf Fluid shear}
\begin{equation}
  \sigma_{\rm t}^2 = \frac{1}{1-v^2}
  \left[\sip^2+\sigm^2-B_1v(2\sip-B1v)
  -\frac{2v\dot{v}}{1-v^2}(\sip-B_1v)+\frac{v^2\dot{v}^2}{(1-v^2)^2}\right] \ .
\end{equation}
The dimensionless shear $\Sigma_{\rm t}$ of the tilted fluid can be
written 
\begin{equation}\label{eq:sit}
  \Sigma_{\rm t}^2 = \frac{\sigma_{\rm t}^2}{\theta_{\rm t}^2}
  = \Sigma_{{\rm t}+}^2 + \Sigma_{{\rm t}-}^2 \ ,
\end{equation}
where
\begin{eqnarray}
  \Sigma_{{\rm t}+}&=&-1+\frac{3(1+\Sip-vA)\left[1-(\gamt-1)v^2\right]}
  {3-2vA+(2\Sip-1)v^2} ,\\
  \Sigma_{{\rm t}-}&=&\frac{3\left[1-(\gamt-1)v^2\right]\Sim}
  {3-2vA+(2\Sip-1)v^2} , \label{eq:Sigma-fluid}
\end{eqnarray}
(see Eqs.\ (A3), (A4) and (A6) in \cite{art:HewittWainwright1992}). 
Note that when both fluids are orthogonal ($v=0$), Eq.\ (\ref{eq:sit})
reduces to
\begin{equation}\label{eq:sito}
  \Sigma_{\rm t}^2=\Sip^2 + \Sim^2 .
\end{equation}

\section{The variable $\chi$}\label{app:chi}

In their study of two orthogonal fluids, Coley \& Wainwright
\cite{art:ColeyWainwright1992} introduced the following variable
instead of $\Omega_0$:
\begin{equation}
  \chi := \frac{\muo - \mut}{\muo + \mut}
  = \frac{\Omo - \Omt}{\Omo + \Omt}\ .
\end{equation}
The evolution equation for $\chi$ becomes
\begin{equation}
  \chi' = -\frac{\left(1-\chi^2\right)}{1+(\gamt-1)v^2}\left\{3\gamo
  \left[1+(\gamt-1)v^2\right]-\gamt \left[3+v^2 - 2v(A+v\Sip)\right]
  \right\} \ .
\end{equation}
For the submanifold $v=0$ corresponding to two orthogonal fluids, the
above equation simplifies to
\begin{equation}
  \chi' = -\tfrac12(1-\chi^2)(\gamo - \gamt)\ .
\end{equation}


\begin{thebibliography}{10}

  \bibitem{Godel}
  K. G\"odel, Rev. Mod. Phys. {\bf 21}, 447 (1949).

  \bibitem{Taub}
  A.~H. Taub, Ann. Math. {\bf 53}, 472 (1951).
  
  \bibitem{art:CollinsHawking1973a}
  C.~B. Collins and S.~W. Hawking, Mon. Not. Roy. Astr. Soc. {\bf 162},  307
  (1973).

  \bibitem{book:WainwrightEllis1997}
  J. Wainwright and G.~F.~R. Ellis, {\em Dynamical systems in cosmology}
  (Cambridge University Press, Cambridge, 1997).

  \bibitem{art:CollinsEllis1979}
  C.~B. Collins and G.~F.~R. Ellis, Phys. Rep. {\bf 56},  65  (1979).

  \bibitem{art:HewittWainwright1992}
  C.~G. Hewitt and J. Wainwright, Phys. Rev. D {\bf 46},  4242  (1992).

  \bibitem{thesis:Harnett1996}
  D. Harnett, Master's thesis, University of Waterloo, 1996.

  \bibitem{art:ColeyWainwright1992}
  A.~A. Coley and J. Wainwright, Class. Quantum Grav. {\bf 9},  651  (1992).

  \bibitem{art:KristianSachs1966}
  J. Kristian and R.~K. Sachs, Astrophys. J. {\bf 143},  379  (1966).

  \bibitem{art:GoliathEllis1999}
  M. Goliath and G.~F.~R. Ellis, Phys. Rev. D {\bf 60},  023502  (1999).

  \bibitem{art:Wainwrightetal1999}
  J. Wainwright, M.~J. Hancock, and C. Uggla, Class. Quantum Grav. {\bf 16},
  2577  (1999).

  \bibitem{art:EllisMacCallum1969}
  G.~F.~R. Ellis and M.~A.~H. MacCallum, Commun. Math. Phys. {\bf 12},  108
  (1969).

  \bibitem{book:ExactSolutions1980}
  D. Kramer, H. Stephani, M.~A.~H. MacCallum, and E. Herlt, {\em Exact
  solutions of Einstein's field equations} (Cambridge University
  Press, Cambridge, 1980). 

  \bibitem{art:Wald1983}
  R.~M. Wald, Phys. Rev. D {\bf 28},  2118  (1983).

  \bibitem{art:RaychaudhuriModak1988}
  A.~K. Raychaudhuri and B. Modak, Class. Quantum Grav. {\bf 5},  225  (1988).

\end{thebibliography}
\end{document}